# Loss-induced high-density multi-mode integrated waveguides arrays

Yanxian Wei[1], Hailong Zhou[1,*], Yunhong Ding[2,3], Zihao Cheng[4,5], Dongmei Huang[4,5], P. K. A. Wai[6], Jianji Dong[1,*], Xinliang Zhang[1,*]

[1]*Wuhan National Laboratory for Optoelectronics, School of Optical and Electronic Information, Huazhong University of Science and Technology, Wuhan 430074, China*

[2]*Department of Photonics Engineering, Technical University of Denmark, Lyngby, Denmark*

[3]*SiPhotonIC ApS, Virum Stationsvej 207, 2830 Virum, Denmark*

[4]*The Hong Kong Polytechnic University Shenzhen Research Institute, Shenzhen 518057, China*

[5]*Photonics Research Institute, Department of Electrical Engineering, The Hong Kong Polytechnic University, Hong Kong, 999077, China*

[6]*Department of Physics, Hong Kong Baptist University, Kowloon Tong, Hong Kong, 999077, China*

*Corresponding author: hailongzhou@hust.edu.cn; jjdong@hust.edu.cn

**Abstract:** The integration density of photonic integrated circuits has been limited by light coupling between waveguides. Traditional approaches to layout the waveguide with high density are based on refractive index engineering to suppress the light coupling between waveguides. However, these methods mostly require sophisticated and sensitive structure design, thus lack universality. Herein, we propose high-density multi-mode-multi-core integrated photonic waveguides by inserting high-loss metal strips between waveguides. We have achieved a 10-spatial-channel multi-mode-multi-core waveguide with total occupying spacing of 6.6 μm. The multi-mode waveguides have a close spacing of 400 nm. The proposed scheme has high fabrication tolerance, ultra-large bandwidth and good compatibility to the complementary metal-oxide-semiconductor technology. It can be applied to any integration platform, any working waveband and any operating mode, providing a universal solution for high-density photonic circuits.

## Introduction

Photonic integrated circuits (PICs), benefiting from their significant advantages of low loss, large bandwidth, interference immunity and compactness, are the leading candidate for optical communications and signal processing[1-3]. The integration density of PICs however is still low compared to electronic circuits because of the large footprint and spacing of photonic waveguides. Great efforts have been made to miniaturize the footprint of photonic components, including the application of plasmonic[4-8] and subwavelength gratings[9]. In particular, the inverse-design methods can reduce the footprint of photonic components by an order of magnitude[10-14]. Reducing the spacing between the waveguides in the photonic circuits is another effective means to improve the integration density by having a more compact layout. In general, the gap spacing of photonic circuits is limited by light coupling between waveguides and should be typically larger than 1.5 μm[15,16]. Refractive index engineering has been proposed to suppress light coupling in compact waveguide arrays, such as inverse-design waveguide array[10], waveguide superlattices[17,18], exceptional coupling metamaterials[19], and dispersion manipulation[20]. These traditional methods however require sophisticated structure design and are sensitive to the fabrication tolerance and/or operating wavelength, and do not support multi-mode operation. A universal solution for high-density integrated photonic waveguides remains in need.

In this work, we first present a high-density multi-mode-multi-core waveguide array based on loss engineering rather than refractive index engineering. The intrinsic high loss of metal materials significantly reduces the light coupling between waveguides with negligible additional loss. We have achieved a 10-spatial-channel multi-mode-multi-core waveguide with total occupying spacing of 6.6 μm. The multi-mode waveguides have a close spacing of 400 nm. Our scheme reveals excellent performances, such as multi-mode operation, large operating bandwidth and large fabrication tolerance. Our work provides a universal solution for crosstalk suppression and high-density integrated photonic waveguides, probably applied to future large-scale on-chip optical systems.

## Results
### Principle

Figures 1a and 1b present the schematic diagrams of a high-density integrated waveguide array without and with high-loss materials between waveguides, respectively. Generally, if the waveguides are placed too close to each other, the light in one waveguide will be coupled to other waveguides, resulting in crosstalk as shown in Fig. 1a. However, by inserting high-loss materials between these silicon waveguides, the coupling can be effectively suppressed, as shown in Fig. 1b. To investigate the light coupling between waveguides, we use the coupled mode theory in a simplified two-waveguide coupled system shown in Fig. 1c, which can be expanded to a multi-waveguide coupled structure. The coupled mode equation of the two coupled waveguide system is given by

$$H\vec{E} = \lambda \vec{E}, \qquad H = \begin{bmatrix} \beta_1 - i\gamma_1 & \kappa \\ \kappa & \beta_2 - i\gamma_2 \end{bmatrix}, \qquad \vec{E} = \begin{bmatrix} E_1 \\ E_2 \end{bmatrix} \qquad (1)$$

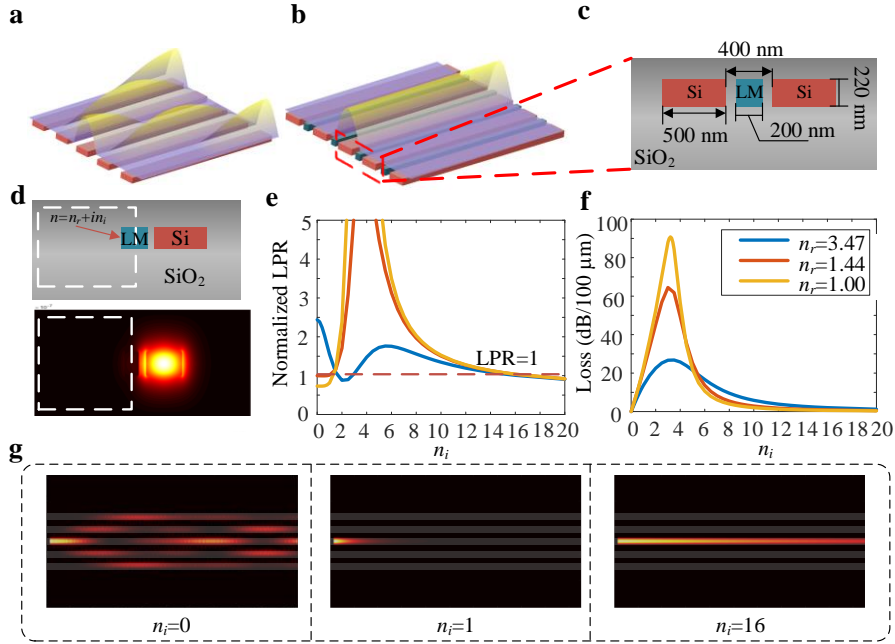

**Fig. 1. Principle of the proposed high-density integrated waveguides. a** High-density integrated waveguides without high-loss materials. **b** High-density integrated waveguides with high-loss materials between the waveguides. **c** Simplified model of a two-waveguide coupled system. **d** Determination of the mode leakage power ratio (LPR). **e** Normalized leakage power ratio of the mode versus $n_i$. **f** The corresponding mode loss. **g** The propagating fields for different $n_i$. LM: loss material.

where $\beta_1$ and $\beta_2$ are the propagation constants of two coupled modes of two adjacent waveguides, $\gamma_1$ and $\gamma_2$ are the loss of two modes, and $\kappa$ is the coupling coefficient. In our previous work[21], we reduced the crosstalk between the two waveguides by increasing the difference in the loss of the two waveguides. Here, we insert high-loss materials between these silicon waveguides to reduce the coupling strength. Since the coupling coefficient $\kappa$ describes the coupling strength between the waveguides, manipulating $\kappa$ has been a effective means to control the coupling dispersion[22,23], and further reducing $\kappa$ is a simple way to suppress crosstalk between the waveguides. In coupled mode theory, $\kappa$ is usually described as the overlapping between the corresponding modes[24], which is the result of the mode leakage[25]. To reduce mode leakage, we put a strip with high material loss between the silicon waveguides. Let the complex refractive index of the strip be $n = n_r + n_i$ where $n_r$ and $n_i$ are the real and imaginary part of $n$, respectively. We use the finite difference time domain (FDTD) method to analyze the mode properties for different $n_r$ and $n_i$. Considering the symmetry of structure, it is only necessary to analyze the eigenmode of one waveguide. Fig. 1d shows the structure of a single waveguide with a high-loss strip material adjoining and the corresponding eigenmode profile. In the simulations, the widths of the silicon waveguides and high-loss strip are 500 nm and 200 nm, respectively. The gap spacing between the silicon waveguides and the height of the silicon waveguide are 400 nm and 220 nm respectively. The high loss of the added strip can effectively suppress the mode leakage. Leakage power ratio (LPR) is defined by the ratio of the power in the white rectangular areas of Fig. 1d and the total power. Thus, LPR represents the power leakage to the adjacent silicon waveguide. We use LPR of structure without a high-loss strip ($n_r = 1.44$, $n_i = 0$) as a reference, and define other LPR with loss strip over reference LPR as normalized LPR. Fig. 1e shows the normalized LPR distribution when $n_i$ varies. The normalized LPR below the red dashed lines indicates that the mode leakage is weakened, otherwise the LPR is enhanced. Fig. 1e shows the LPR of three media, silicon (blue), silica (brown) and air (orange), with refractive indices $n_r = 3.47, 1.44,$ and 1 respectively. From Fig. 1e, for $n_i = 0$, LPR decreases when $n_r$ decreases, because of the increase in the refractive index difference between the core and the cladding. For $n_r = 1.44$ and $n_r = 1$, the LPR first increases and then decreases when $n_i$ continues to increase. If $n_i$ is sufficiently large, the LPRs of all three structures will be lower than that of the reference structure, indicating less overlapping between the two modes in the two-waveguide coupled system.

Inserting high-loss materials between the silicon waveguides is likely to introduce high loss to the modes. However, from Fig. 1f, the optical loss first increases with $n_i$ until $n_i$ reaches a breaking phase point. After the point, the optical loss begins to reduce when $n_i$ increases constantly, which is similar to the behaviors in Non-Hermitian photonic systems[21,26]. Therefore, by using materials with sufficiently high loss, one can achieve both mode leakage suppression and low loss propagation.

We then calculate the propagating field distributions in the waveguide arrays. We assume the arrays consists of five silicon waveguides ($n = 3.47$) with high-loss strips ($n_1 = 3.47 + n_i i$) placed in the gaps between the waveguides. The length of all waveguides is chosen to be 200 μm. The gap between the silicon waveguides is chosen to be 400 nm and the width of the loss strips is 200 nm. Fig. 1g and show the simulation results for different values of $n_i$. When $n_i = 0$, light will oscillate in these waveguides resulting in crosstalk. The coupling between the waveguides however will weaken when $n_i$ increases from zero, but the loss suffered by the light absorption will also increases. When $n_i$ is large sufficiently, light will only propagate in the center waveguide with negligible loss. The same mechanism applied to higher-order modes.

**Experiment**

We choose metals with a high imaginary refractive index as the high-loss strips. From the discussion above, the intrinsic high imaginary refractive index of metal would render the light propagating in the silicon waveguides a low crosstalk and non-negligible loss. We design and fabricate devices with multi-mode waveguides whose cross-section width is 1μm shown in Figs. 2a and 2b, and measure the transmission spectra of each waveguide for input $TE_0$ or $TE_1$ mode. Figs. 2c and 2d show the results at 1550 nm. The crosstalk of the referenced $TE_0$ mode in the multi-mode structure is low since the mode field is far from the edge of waveguide. Nevertheless, adding metal strips can still distinctly increase the extinction ratio between the waveguides. The crosstalk of the referenced $TE_1$ mode is significant. Compared to the reference structure, the crosstalk between the waveguides for $TE_0$/$TE_1$ mode input to the structure with Al strips is significantly suppressed. An extinction ratio of >10 dB is observed. Therefore, we achieve a 10-channel multi-mode-multi-core waveguide with a total occupying spacing of 6.6 μm. The results demonstrated that the proposed design can be applied to higher order modes, and thus enables applications such as multi-core-multi-mode bus waveguides.

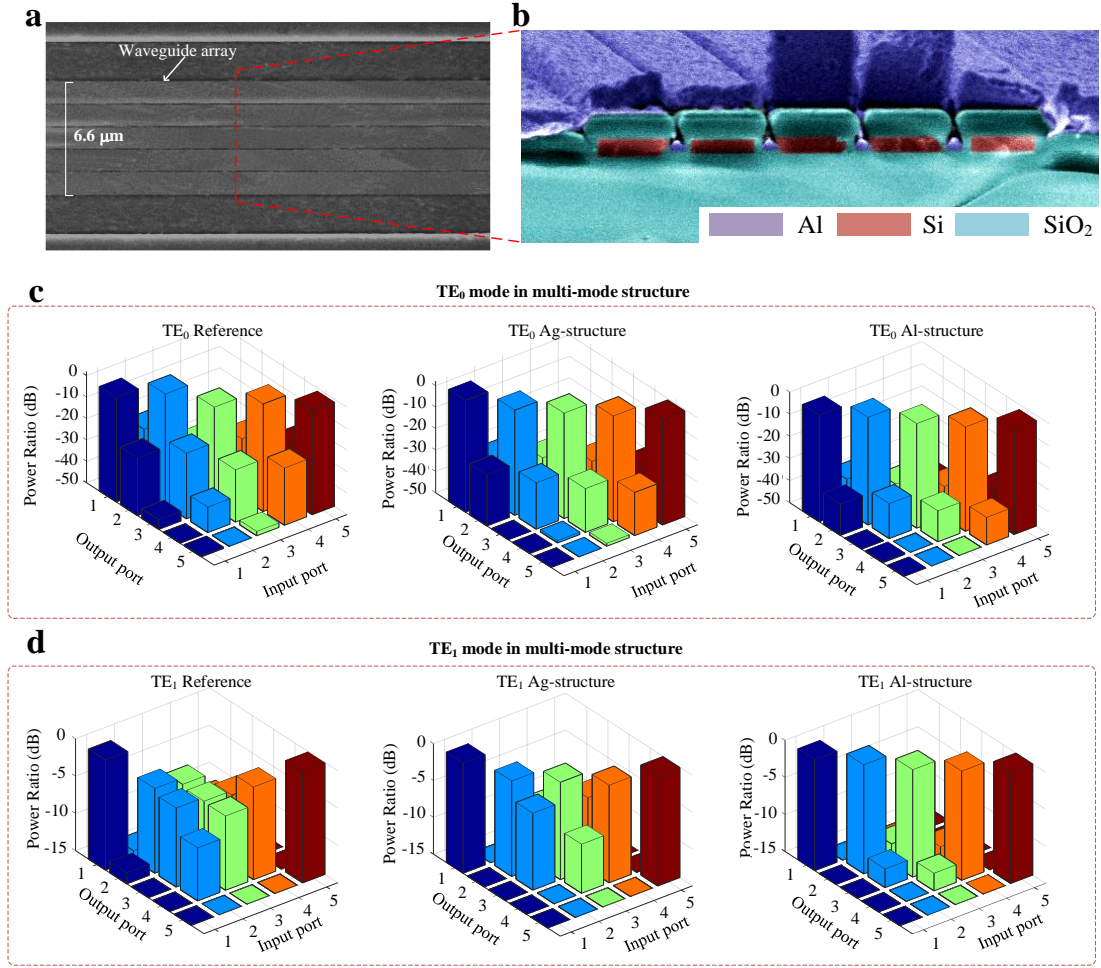

**Fig. 5. Experimental results of multi-mode structure. a** The scanning electron microscope (SEM) pictures of the chip. **b** The cross-sectional view of the coupling region. **c** The measured power ratios of $TE_0$ mode input to the multi-mode structure. **d** The measured power ratios of $TE_1$ mode input to the multi-mode structure.

## Conclusion:

In summary, we have proposed and demonstrated a high-density multi-mode-multi-core waveguide array based on loss engineering. By inserting metal strips into the gaps between silicon waveguides, the crosstalk between the waveguides can be greatly suppressed. We have achieved a 10-channel multi-mode-multi-core waveguide with total occupying spacing of 6.6 μm. The multi-mode waveguides have a close spacing of 400 nm. Compared to refractive index engineering, our scheme reveals excellent performances, such as multi-mode operation, large operating bandwidth and large fabrication tolerance. The proposed device provides a competitive method to increase the integration density of silicon photonics chips.

## Acknowledgements:

This work was partially supported by the National Key Research and Development Project of China (2022YFB2804200), the National Natural Science Foundation of China (U21A20511, 62075075, 62275088), the Innovation Project of Optics Valley Laboratory (Grant No. OVL2021BG001).

## Authors' contributions

Y.X.W. and H.L.Z. conceived the idea. Y.X.W., H.L.Z. Z.H.C. D.M.H. and P. K. A. W. discussed improved the theory. Y.X.W. performed the numerical simulations and designed the device. Y.X.W. performed the experiments. Y.H.D. fabricated the chip. Y.X.W., D.M.H. and P. K. A. W. discussed the results. J.J.D and X.L.Z supervised the study. All authors contributed to the writing of the paper.

## Data availability

All the data related to this paper are available from the corresponding authors upon request.

**Additional information:** There is no additional information for this article.

**Competing financial interests:** The authors declare no competing financial interests.